\documentclass[prl,aps,twocolumn,groupedaddress,showpacs,floatfix]{revtex4}

\usepackage{amsmath,amssymb,multirow,epsfig,bm}

\newcommand{\etal}{{\it et al.,\;}}
\newcommand{\beq}{\begin{equation}}
\newcommand{\eeq}{\end{equation}}
\newcommand{\bea}{\begin{eqnarray}}
\newcommand{\eea}{\end{eqnarray}}

\newcommand{\nn}{\nonumber}
\newcommand{\benn}{\begin{displaymath}}
\newcommand{\eenn}{\end{displaymath}}

\begin{document}

\title{\bf Collective oscillations of a trapped Fermi gas near the
unitary limit }

\author{ Aurel Bulgac$^1$ and George F. Bertsch$^{1,2}$}
\affiliation{$^1$Department of Physics, University of
Washington, Seattle, WA 98195--1560, USA}
\affiliation{$^2$ Institute for Nuclear Theory,
University of Washington, Seattle, WA 98195-1550, USA}

\date{\today}

\begin{abstract}

We calculate the oscillation frequencies of trapped Fermi condensate
with particular emphasis on the equation of state of the interacting
Fermi system.  We confirm Stringari's finding that the frequencies are
independent of the interaction in the unitary limit, and we extend the
theory away from that limit, where the interaction does affect the
frequencies of the compressional modes only.

\end{abstract}



\maketitle


The remarkable advances in producing and measuring properties of
atomic condensates give a strong impetus to develop theory to meet the
challenges of interpreting the experiments.  In the case of fermion
condensates, one is now at the early stages of coming to an
understanding of the first experimental results \cite{thomas,grimm}.
One of the characteristic properties is the frequencies of normal
modes of vibration.  Stringari \cite{stringari} has developed the
theory at the unitary limit, finding that the oscillation frequencies
are independent of the details of the interaction.  Here we extend the
theory away from the unitary limit where the interaction has some
effect.

The "unitary limit" is a term to describe a two-component Fermi gas
with a short-range interaction, characterized by a scattering length
that is large compared to the length scale set by the particle
density.  This limit was discussed in 1999, when one of us (GFB)
formulated as a challenge to many-body theorists to clarify the
structure of the ground state of a fictitious neutron matter,
interacting with an infinite scattering length \cite{george}. At the
time when the challenge has been issued it was not really clear even
if such matter is stable in principle, as either a system of bosons
\cite{efimov} or a system of three or more fermion species \cite{abve}
is known to be unstable, due to the well known Efimov effect. From
simple dimensional arguments it was clear that the energy per particle
of a two fermion species should be proportional to that of the 
free Fermi gas, with a universal constant of proportionality.
This may be expressed as
\beq
\varepsilon \equiv \frac{E}{N}= \frac{3}{5}\frac{\hbar^2k_F^2}{2m} \xi ,
\eeq
where $k_F$ is the Fermi wave vector and $\xi$ some pure number. Even
though estimates of this number have been generated at the time
\cite{baker}, the most convincing argument that this number is
positive indeed came only recently, both in theory \cite{carlson},
namely $\xi \approx 0.44$, and experiments \cite{experiments}.
Experiments indicate so far that $\xi \approx 0.5$ with very large
error bars. The theory also provides strong arguments that such a
fermionic system should be superfluid at zero temperature, while the
experiment is still lacking in this respect. The information inferred
from various experiments so far is, at best, of a rather indirect
nature and no direct unambiguous evidence of superfluidity has been
reported until now.  Very likely, only the direct observation of
vortices will provide the ultimate experimental proof that such
systems can sustain a superflow. There are strong theoretical
arguments that such vortices should be almost as easily visible in the
Fermi dilute atomic clouds \cite{abyy} as in Bose dilute atomic clouds
\cite{bose}.

In the Green Function Monte Carlo calculations of the
Refs. \cite{carlson,chang} it was established that the ground state
energy per particle of the homogeneous phase of a system of two
fermion species, interacting with a very large scattering length (much
larger than the radius of the interaction $r_0$ (namely $|a|\gg r_0$),
in the dilute regime ($nr_0^3\ll 1$, where $n=n_\uparrow +n_\downarrow
= k_F^3/3\pi^2$) is
\bea
\label{energy}
& & \varepsilon(n) =  \frac{3}{5}\frac{\hbar^2k_F^2}{2m}
F\left(\frac{1}{k_Fa}\right ) \nn \\
& &  =  \frac{3}{5}\frac{\hbar^2k_F^2}{2m}
\left [ \xi -\frac{\zeta}{k_Fa} - \frac{5\upsilon}{3k_F^2a^2}
+ {\cal{O}}\left ( \frac{1}{(k_Fa)^3}\right ) \right ],
\eea
where $F(x)$ is a universal function and the constants $\xi \approx
0.44$, $\zeta \approx 1$ and $\upsilon \approx 1$ (the last two values
have been extracted by us from the numerical results provided by the
authors \cite{chang}).  We shall discuss the normal modes of
oscillation for a Fermi system in a harmonic trap, whose energy
per particle is given by Eq. (\ref{energy}).  We note that the last term
is independent of density and can be dropped from the present analysis.

We assume throughout that the system behaves hydrodynamically,
i.e. that the pressure tensor is isotropic. If the system is
superfluid, then as long as the oscillation frequency is below the gap
frequency needed to break-up a Cooper pair, this condition is expected
to be fulfilled.  The ground state density satisfies the equation
\beq
 {\nabla^2}P(n_0) +  {\bf \nabla}[n_0\cdot
{\bf \nabla}U ] = 0, \label{eq:gs}
\eeq
where $n_0$ is the ground state density, $U$ is the trapping
potential, and $P$ is the pressure. $P$ is related to the energy per
particle $\varepsilon(n)$ by
\beq
P(n)= n^2 \frac{d\varepsilon(n)}{dn}.
\eeq
The slow and small-amplitude normal modes satisfy the following equation

\beq
  - m\nabla^2 [c^2(n_0)n_1]
-{\bf \nabla}(n_1\cdot {\bf \nabla}U)=m \omega^2 n_1 \label{eq:osc}
\eeq
where $n_1$  is the oscillating density and
\beq
c^2(n) = \frac{1}{m}\frac{d P(n) }{  dn}
\eeq
gives the speed of sound $c$ in a uniform condensate at density $n$.
It is straightforward to show that Kohn's generalized theorem
\cite{kohn} (stating that the frequencies of the dipole modes are the
trap frequencies) is satisfied.  It is convenient to make a change of
variable in this equation, replacing $n_1$ by the variable $f_1$
defined by
\beq
n_1 = n_0 f_1 / c^2(n_0).
\eeq
After some algebra making use of Eq. (\ref{eq:gs}) we find
\beq
\label{eom2}
\nabla \cdot (n_0 \nabla f_1) = - \omega^2 \frac{n_0}{c^2(n_0)} f_1.
\eeq
This equation has the formal advantage of being Hermitian, and
thus easier to study in perturbation theory.

As noted before \cite{menotti}, the equations of motion admit simple
scaling solutions for gases in harmonic traps that satisfy polytropic
equations of state.  Before discussing a perturbative treatment, we
shall present results of an analysis under the assumption of a
polytropic equation of state.  As we were completing this work, we
learned of a similar analysis by Heiselberg, who presents a general
solution in terms of hypergeometric functions (which reduce to polynomials 
in this case) \cite{henning}. The polytropic equation of state is
\beq
P(n) = \alpha n^\gamma ,
\eeq
where the constant $\gamma$ is the adiabatic index of the gas.
The ground state density is given by
\beq
\label{n0}
n_0({\bf r}) = \left[\frac{\gamma -1}{\gamma \alpha }
\mu({\bf r}) \right ] ^\nu.
\eeq
Here we introduced
\beq
\mu({\bf r}) = \mu_0 -U({\bf r}), \quad  \gamma = 1 +\frac{1}{\nu},
\eeq
where $\mu_0$ is the chemical potential and the harmonic trap potential
is given by
\beq
U({\bf r}) = \frac{m \omega_0^2 (x^2+y^2+\lambda^2 z^2)}{2}.
\eeq
The corresponding local sound speed is given by
\beq
\label{c2}
c^2({\bf r} ) = \frac{ \mu({\bf r}) }{\nu m}.
\eeq
Using these expressions, Eq. (\ref{eom2}) can be written in the form
\beq
\label{eom:f1}
 \frac{\mu({\bf r})}{\nu m} \Delta f_1
+ \nabla \frac{\mu({\bf r})}{m}\cdot  \nabla f_1 = -\omega^2 f_1.
\eeq
Due to the particular polynomial form of $\mu(r)$ the eigenfunctions
$f_1$ have a polynomial character as well. From Eq. (\ref{energy}) one
can extract an effective adiabatic index for a Fermi gas in the
vicinity of a Feshbach resonance
\beq \label{gamma}
\gamma =\frac {d \ln P}{d \ln n}=
\frac{5}{3}\left [ 1 + \frac{\zeta}{10 \xi k_Fa } +
{\cal{O}}\left ( \frac{1}{(k_Fa)^2} \right ) \right ]
\eeq

Note that the quadrupole mode in the spherical condensate and the
transverse quadrupole modes in the deformed condensate do not depend
on the equation of state.  That is because the flow in these modes is
incompressible and the internal energy does not change during the
oscillation cycle.  The frequencies of the monopole and of the two
compressional modes in the deformed condensate have a dependence on
$\gamma$, and we can use that to estimate the frequency shift.

Eq. (\ref{gamma}) shows that the effective adiabatic index is larger
than 5/3 on the BEC side of the Feshbach resonance (when $a>0$). This
behavior implies that the frequency of the radial oscillations should
increase as well when going from the BCS to the BEC side of the
Feshbach resonance.  This conclusion agrees with the conjecture made
by Stringari \cite{stringari}, except that now the frequency shift has
been evaluated explicitly in terms of the properties of the system. In
the unitary limit, when $\gamma = 5/3$, these results also agree with
previous results for the non-interacting Fermi systems in traps
\cite{amoruso} and the results for a superfluid Fermi system away from
a Feshbach resonance in a spherical trap \cite{baranov}.  The
quadrupole frequencies obtained using scaling solutions \cite{menotti}
and the sum-rule approach \cite{vichi}, in the limit of a
non-interacting Fermi gas, are different, and indeed correspond to the
diabatic limit or collisionless regime \cite{diabatic}. In this limit
the sphericity of the Fermi surface is lost during oscillations and
the cloud behaves like a normal Fermi gas in Landau's zero sound
regime.

\begin{table}

\caption{Results for a polytropic gas. For $\lambda \ll 1$ only leading
terms
are shown and $c_{1,2,3}$ are some constants. }

\begin{tabular}{|cc|c|c|}
\colrule
trap type     &   mode & $f_1$             &   $\omega^2/\omega_0^2$  \\

\colrule
spherical     &  L = 1 &  $x,y,z$  & 1      \\
$\lambda = 1$ &  L = 2 &  $xy$, etc.  &$ 2$     \\
              &  L = 0 &  $1-2  r^2$ &  $3\gamma - 1 $    \\
\colrule
axial &  $M =\pm 2$ & $xy, \;x^2-y^2$  & $2$ \\
$ \lambda\ll 1$    &   $M = \pm 1$    &   $xz$, $yz$ & $1+\lambda^2$ \\
     &   radial   & $x^2+y^2+c_1\lambda^2z^2+c_2$  & $2\gamma $ \\
        &  axial  & $z^2 +c_3$  & $\lambda^2(3\gamma -1)/\gamma $ \\
\colrule
\end{tabular}
\end{table}

The polytropic analysis is useful to show the basic dependence on the
system parameters, but the parameter $k_F$ is ill-determined, due to
the non-uniformity of the condensate.  We shall therefore use
perturbation theory to make a more quantitative analysis. Since
Eq. (\ref{eom2}) is Hermitian, the following formula is variational in
the $f_1$,
\beq
\omega^2 = \frac{\int |\nabla f_1|^2 n_0 d^3 r
}{\int f_1^2 \frac{n_0}{c^2} d^3 r}.              \label{omega2}
\eeq
This implies that one can ignore the perturbation in $f_1$ in calculating
the first-order shift in the frequency.  That shift is then given by
\beq
\frac{\delta \omega^2}{\omega^2} =
\frac{\int |\nabla f_1|^2 \delta n d^3 r}
{\int |\nabla f_1|^2 n_0 d^3 r}
-\frac{\int f_1^2 \delta \left ( \frac{n}{c^2} \right ) d^3 r}
{\int f_1^2  \frac{n_0}{c^2}  d^3 r}.
\label{domega2}
\eeq
To apply this formula, we take the unperturbed densities
$n_0$ and $n_0/c^2$ from Eq. (10) and (13), setting $\gamma=5/3$.
The needed functions are
\beq
n_0 = \frac{1}{3 \pi^2}\left(\frac{2 m \mu }{\xi \hbar^2}\right)^{3/2},
\quad c^2 = \frac{2\mu}{3 m}.
\eeq

We now have to determine the change in $n$ and $n/c^2$ induced by
the second term in the energy function Eq. (\ref{energy}).  Including
that term, the chemical potential satisfies the equation
\beq
\mu_0  =\varepsilon + n\frac{d \varepsilon } {d n} + U =
\frac{\hbar^2k_F^2}{2m}\xi -\frac{2\hbar^2 k_F}{5ma}\zeta + U.
\label{eq:mu}
\eeq
We may use this expression to evaluate the first-order change in
ground state number density $n_0$.  Since the frequency does not
depend on $\mu_0$, we may hold it fixed in doing the variation. The
result is
\beq
\delta n = \frac{4}{5\pi^2} \frac{m \zeta}{ \hbar^2 \xi^2 a} \mu
\eeq
In the same way we include the perturbation in the formula for $c^2$
and obtain that the corresponding first order change in $n/c^2$ is
given by
\beq
\delta \left ( \frac{n}{c^2} \right )
= \frac{4}{5\pi^2}  \frac{m^2\zeta }{\hbar^2 \xi^2 a}
\eeq
We next evaluate the integrals in Eq. (\ref{domega2}). It is convenient
to change lengths to a dimensionless form, scaling them by the
transverse condensate radius $R=\sqrt{2 \mu_0/m\omega_0^2}$,
\beq
R (\tilde x, \tilde y,\tilde z) = (x,y,\lambda z).
\eeq
In terms of the scaled variable $\tilde \mu = 1 - \tilde r^2$,
Eq. (\ref{omega2}) for the unperturbed frequency may be expressed
\beq
\omega^2 = \frac{\omega_0^2}{3}\frac{\int |\tilde \nabla  f_1|^2
\tilde \mu^{3/2} d^3\tilde r}{\int f_1^2 \tilde \mu^{1/2}d^3\tilde r}.
\eeq
Eq. (12) for the frequency shift becomes
\beq
\label{K}
\frac{\delta \omega^2}{\omega^2} =
\frac{\zeta}{\xi^{1/2}}
\frac{\hbar}{m  \omega_0 R a } K
=  \frac{\zeta}{\xi }\frac{1 }{k_F(0)a} K,
\eeq
where the dimensionless factor $K$ is given by
\beq
\label{k}
K = \frac{6\int|\nabla f_1|^2 \tilde \mu d^3 \tilde r}
{5\int|\nabla f_1|^2 \tilde \mu^{3/2} d^3 \tilde r}
- \frac{4\int f_1^2 d^3 \tilde r}
{5\int f_1^2 \tilde \mu^{1/2} d^3 \tilde r}
\eeq
and $k_F(0)$ is the value of the local Fermi momentum at the center of
the trap.  The prefactor in Eq. (\ref{K}) displays the scaling of the
frequency shift with respect to the physical parameters of the
condensate.  As expected, the shift is inversely proportional to the
combination  $k_Fa$. Finally, it has a nontrivial dependence on
$\xi$ and $\zeta$, the universal parameters defining the energy per
particle in the vicinity of a Feshbach resonance.

All the needed radial integrals have the form
\bea
I_{m,n} &=& \int^1_0 \tilde r^m (1-\tilde r^2)^n d \tilde r, \\
J_{m,n} &=& \int^1_0 \tilde r^m (1-\tilde r^2)^n
(1 - b\tilde r^2+c\tilde r ^4) d \tilde r,
\eea
after integration over angular variables. Since in Eq. (\ref{k}) both
denominators and numerators have the same angular dependence, the
specific values of the angular integrals cancels out in the case of
incompressible modes. We present the results for the various cases in
Table II.  One sees that the shift of the dipole mode vanishes, as
required by the generalized Kohn's theorem. The shift also vanishes
for the pure quadrupole modes, for reasons noted earlier.  The cases
of most interest are the monopole mode in spherical traps ($\lambda
=1$) and the $M=0$ modes in axially deformed (essentially cylindrical)
traps with $\lambda \ll 1$.  The $K$ factors are non-vanishing in
these cases, but they are rather small, for example, $K_{radial}
\approx 0.12$.  This has the same order of magnitude as the factor
$1/10$ in Eq. (15) for the change in the effective adiabatic index.
We also note that the sum of factors $K$ determining the shifts for
the radial and axial modes equals the factor $K$ for the pure monopole
mode in the spherical case.


\begin{table}
\caption{ Results for $K$.}
\begin{tabular}{|c|c|ccc|}
\colrule
trap type    &   mode   & $f_1$ &   $\omega$    &  $K$\\
\colrule
spherical    &   dipole  &  $ z$ & $\omega_0$ & 0 \\
$\lambda=1$  &  monopole &  $1-2 r^2$&  $2\omega_0$ &
$\frac{256}{525\pi}$  \\
             &   quadrupole & $ x y$ &$\sqrt{2}\omega_0$ & 0    \\
\colrule
axial & $M= \pm 2$ & $xy,\;x^2-y^2$  &$\sqrt{2}\omega_0$ & 0\\
$\lambda\ll 1$& $M= \pm 1$ & $xz$, $yz$  &$\omega_0$ & 0\\
 &radial & $ x^2+y^2+\frac{2}{5}\lambda^2z^2 -\frac{2}{5}$
& $\sqrt{\frac{10}{3}}\omega_0$ & $\frac{1024}{2625\pi}$ \\
&axial & $1-6\lambda^2 z^2$ & $\sqrt{\frac{12}{5}}\lambda \omega_0$
&$\frac{256}{2625\pi}$ \\
\colrule
\end{tabular}
\end{table}

The two experimental results available so far \cite{thomas,grimm} are
still in noticeable disagreement with each other to permit a detailed
comparison with theory. Nevertheless, both experiments show distinctly
a qualitative agreement with theory as far as the character of the
frequency shift is concerned, in the vicinity of the Feshbach
resonance.  The fact that both experiments seem to favor the adiabatic
character of the oscillations should not be interpreted yet as a
confirmation of the existence of superfluidity in these systems, since
the sphericity of the Fermi surface can be maintained by
collisions. In this respect the situation here is to some extent
similar to the expansion of a cold Fermi gas \cite{duke,mit}. It is
also important to determine experimentally the frequencies of the
transversal quadrupole modes, since a shift in these frequencies can
point to a complex structure of the cloud, similar to that discussed
in Ref. \cite{at_mol}.

This work was supported in part by the Department of Energy under
grants DE-FG06-90ER40561 and DE-FG03-97ER41014. We thank R. Grimm and
C. Chin for initial discussions and correspondence leading us to this
study and J. Carlson, S.Y. Chang and V.R. Pandharipande for sharing
with us their results prior to publication.

{\it Note added.} After submitting this manuscript we learned of a few
another studies of Fermi systems using the polytropic equation of
state \cite{hu}.



\begin{thebibliography}{99}


\bibitem{thomas} J. Kinast \etal Phys. Rev. Lett. {\bf 92}, 150402
(2004).

\bibitem{grimm} M. Bartenstein \etal cond-mat/0403716.

\bibitem{stringari} S. Stringari, Europhys. Lett. {\bf 65}, 749
(2004).

\bibitem{george} G.F. Bertsch, {\it Many-Body challange problem}, see
R.A. Bishop, Int. J. Mod. Phys. {\bf B 15}, {\it iii}, (2001).

\bibitem{efimov} V. Efimov, Phys. Lett. {\bf B33}, 563 (1973);
Sov. J. Nucl. Phys. {\bf 12}, 589 (1971); Nucl. Phys. {\bf A210}, 157
(1973); Nucl. Phys. {\bf A362}, 45 (1981).

\bibitem{abve} A. Bulgac and V. Efimov, Sov. J. Nucl. Phys. {\bf 22},
296 (1975).

\bibitem{baker} G.A. Baker, Jr., Int. J. Mod. Phys. {\bf B 15}, 1314
  (2001); H. Heiselberg, Phys. Rev. A {\bf 63}, 043606 (2001).

\bibitem{carlson} J. Carlson, \etal Phys. Rev. Lett. {\bf 91}, 050401
  (2003).

\bibitem{experiments} K.M.O'Hara \etal Science {\bf 298}, 2179 (2002);
  C. A. Regal \etal Nature {\bf 424}, 47 (2003), see also
  K.E. Strecker \etal Phys. Rev. Lett. {\bf 91}, 080406 (2003);
  J. Cubizolles \etal Phys. Rev. Lett. {\bf 91}, 240401 (2003);
  S. Jochim \etal Phys. Rev. Lett.  {\bf 91}, 240402 (2003);
  M. Greiner \etal Nature {\bf 426}, 537 (2003); M.W. Zwierlein \etal
  Phys. Rev. Lett. {\bf 91}, 250401 (2003); S. Jochim \etal Science
  {\bf 302}, 2101 (2003).

\bibitem{abyy} A. Bulgac and Y. Yu, Phys. Rev. Lett. {\bf 91}, 190404
  (2003).

\bibitem{bose}K.W. Madison, F. Chevy, J. Mod. Opt. {\bf 47}, 2715
  (2000); F. Chevy, \etal Phys. Rev. Lett. {\bf 85}, 2223
  (2000);J.R. Abo--Shaeer, \etal Science {\bf 292}, 476 (2001);
  E. A. Cornell and C. E. Wieman, Rev. Mod. Phys. {\bf 74}, 875
  (2002); W. Ketterle, Rev. Mod. Phys. {\bf 74}, 1131 (2002).

\bibitem{chang} See the talk by S.Y. Chang at the INT Workshop on
 Strongly Interacting Fermi Systems, November 18-20, 2003, posted at
 {\tt http://int.phys.washington.edu/talk\_list.html} and S.Y. Chang
 \etal physics/0404115.

\bibitem{kohn} J.F. Dobson, Phys. Rev. Lett. {\bf 73}, 1838 (1997).

\bibitem{menotti} C. Menotti \etal Phys. Rev. Lett. {\bf 89}, 250402
(2002).

\bibitem{henning}  H. Heiselberg, cond-mat/0403041.

\bibitem{amoruso} M. Amoruso, \etal Eur. Phys. J.  {\bf D 7}, 441
  (1999).

\bibitem{baranov} M.A. Baranov and D.S. Petrov, Phys. Rev. A {\bf 62},
  041601(R) (2000).

\bibitem{vichi} L. Vichi and S. Stringari, Phys. Rev. A {\bf 60}, 4734
(1999).

\bibitem{diabatic}  G.F. Bertsch \etal physics/0403125.

\bibitem{duke} K.M. O'Hara \etal Science {\bf 298}, 2179 (2002).

\bibitem{mit} S. Gupta \etal Phys. Rev. Lett. {\bf 92}, 100401 (2003).

\bibitem{at_mol} A. Bulgac, cond-mat/0309358.

\bibitem{hu} M. Cozzini and S. Stringari, Phys. Rev. Lett. {\bf 91},
070401 (2003); Y.E. Kim and A.L. Zubarev, cond-mat/0403085; H. Hu,
\etal cond-mat/0404012.

\end{thebibliography}
\end{document}